\RequirePackage{fix-cm}
\documentclass[twocolumn,epjc3]{svjour3}  
\smartqed  
\RequirePackage{graphicx}
\usepackage{hyperref}
\hypersetup{colorlinks,citecolor=blue}

\usepackage{color}
\usepackage{amsmath}
\usepackage{amssymb}
\journalname{Eur. Phys. J. C}
\begin{document}

\title{Quark stars in 4-dimensional Einstein-Gauss-Bonnet gravity}


\author{Ksh. Newton Singh\thanksref{e1,addr1}
        \and
        S. K. Maurya \thanksref{e5,addr2} 
        \and
        Abhisek Dutta\thanksref{e4,addr3}
         \and
        Farook Rahaman\thanksref{e2,addr4} 
        \and
        Somi Aktar\thanksref{e3,addr4} 
}

\thankstext{e1}{e-mail: ntnphy@gmil.com}
\thankstext{e5}{e-mail: sunil@unizwa.edu.om}
\thankstext{e4}{e-mail: abhisek96.physics@gmail.com}
\thankstext{e2}{e-mail: rahaman@associates.iucaa.in}
\thankstext{e3}{e-mail: somiaktar9@gmail.com}


\institute{Department of Physics, National Defence Academy, Khadakwasla, Pune-411023, India. \label{addr1}
\and
Department of Mathematical and Physical Sciences, College of Arts and Science, University of Nizwa, Sultanate of Oman. \label{addr2}
\and
Department of Physics, Jadavpur University, Kolkata 700032, India. \label{addr3}
\and
Department of Mathematics, Jadavpur University, Kolkata 700032, India. \label{addr4}
          }

\date{Received: date / Accepted: date}

\maketitle

\begin{abstract}
The article explores the first exact solution in a four-dimensional EGB-gravity with an anisotropic matter source. The solution has been found by two assumptions, a metric potential and the MIT-bag equation of state. Further, the solution has been tested with several physical constraints. Finally, using the boundary condition we have also estimated the range of bag constant $\mathcal{B}$ by varying the mass of the structure and the Gauss-Bonnet coupling constant $\alpha$. This allows us to check the physical viability of the solution by comparing it with the existing accepted range of bag constant $\mathcal{B}$.

\end{abstract}

\section{Introduction}\label{sec1}
A well established fact among the scientific society is the finest methodology to explain phenomena occurring in 4-dimensional spacetime is the Einstein General Relativity which is pertinently governed by the Einstein Hilbert action ($S_{EH}$) and nevertheless the cosmological constant ($\Lambda_0$), can be represented as -
$$S_{EH} = \int d^D x \sqrt{-g} \left[\frac{M_P^2R}{2} - \Lambda_0 \right]$$
Here $M_P$ represents planck's mass and $D=4$ i.e. for $4-D$ spacetime. The planck's mass is used in correspondence to express to the gravitational coupling strength. In the realm of higher dimensional spacetime i.e. $D>4$, \cite{ghosh2021phase} performing quadratic curvature corrections to the Einstein Hilbert action leads to the Einstein$-$Gauss$-$ Bonnet (EGB) theory. Lovelock's Theorems \cite{lovelock1971einstein,lovelock1972four} into higher dimensions are fundamentally generalisation of Einstein's Equation in higher dimensions. The main advantage in Lovelock's theorem is the equations of motions are second order variants of the metric which makes it much more mathematically lucid with respect to the other gravity theories that are 4th or higher derivatives of the metric. It is an intriguing matter that the \cite{lanczos1938remarkable} Einstein Gauss Bonnet gravity can be classified as a special case of Lovelock's theorem, as the quadratic curvature corrections obtained in this method is through variation of the Gauss - Bonnet term. Zwiebach \cite{zwiebach1985curvature} provided us that Gauss-Bonnet term is the lone pertinent method to perform higher derivative combination for the leading correction to Einstein theory, in the low energy effective string theory. The EGB theory basically amalgamates a single term to Einstein-Hilbert action which is  quadratic in Riemann tensor which naturally serves as the action of Low energy heterotic string theory \cite{wiltshire1986spherically,wheeler1986symmetric,boulware1985string} . The low energy effective string theory is quite a promising theory as it appears to be a prominent theory of gravity in higher dimension as it complies higher order curvature terms in the action. This theory is considered be a contender to remove the singularity problem in Einstein's general relativity.

Recently, Glavan and Lin \cite{glavan2020einstein} introduced a general covariant modified version of the theory which sorts out multiple non-trivial problems occurring in $4-D$ spacetime and higher dimensions. The Gauss-Bonnet term ($\alpha$) in $4-D$ spacetime is nothing but Euler Density and hence a total derivative and does not effect the gravitational dynamics. But on higher dimensions, $D>4$, it contains an exterior result of the total derivative along with a $(D-4)$ form. In the higher dimension scenario, Mardones et al. \cite{mardones1991lovelock} and Torii et al. \cite{torii2008n+} expressed that the contribution of the Gauss-Bonnet term is proportional to a vanishing factor of $(D-4)$ . Glavan and Lin \cite{glavan2020einstein} rescaled the coupling factor i.e. $\alpha$ to $\frac{\alpha}{(D-4)}$. This rescaling is very much congruous to general relativity as it maintains equal degrees of freedom in all dimensions and also rules out the Ostrogradsky instability  \cite{woodard2015theorem}. The regularization of the  Gauss-Bonnet term at $D=4$ providing non-trivial contribution in dynamics was first shown by Tomozawa \cite{tomozawa2011quantum}.\\
This modified gravity theory has versatile applications and hence it appeared to be quite intriguing to researchers in the field of Astrophysics and Cosmology. Glavan and Lin \cite{glavan2020einstein} themselves applied the theory to the case of static spherically symmetric vacuum black hole which fetches out quite a different results from what was obtained for Schwarzchild Black Holes and is open to debate in terms of acceptability. Gurses et. al \cite{gurses2020there} showed that, Einstein Gauss Bonnet gravity does not provides any intrinsic 4D solution in terms of metric, though the viability of the solutions is appropriate in $D>4$ scenarios. Further extension of this theory was applied to evaluate the thermodyanmics of Rotating and non-rotating black holes and calculate parameters like mass, temperature, entropy etc using EGB gravity. \cite{ghosh2020generating,konoplya2020stability,kumar2020hayward,kumar2020bardeen,zhang2020superradiance}. Along with that, this theory is successfully applied for the cases of Black gravitational Lensing of both strong and weak types \cite{ghosh2020generating}, thermodynamics of EGB Anti de Sitter Black Holes \cite{mansoori2021thermodynamic}, evaluating the Quasi Normal Modes \cite{mishra2020quasinormal}, Geodesic motions and for spinning test particles \cite{zhang2020spinning}, Hawking radiation \cite{zhang2020greybody} and even in wormhole solutions \cite{kimet2020wormholes}.
Apart from the Black Hole problems, EGB find its application on case of Neutron Stars. The core of the neutron stars is still an area not much explored yet. Recent inventions of Neutron stars of 2$M_\odot$ pushes the equation of state of matter inside the stars to farthest extent but the inner sanctum remains a mystery. The determination of inner core matter is quite non-trivial as the composition varies with the nature of interaction and quark matter, hyperon matter, Bose Einstein condensate strange mesons are all considered to be the core matter or might be present in fraction. But the the invention of 2$M_\odot$ Neutron stars like PSR J1614-2230 [$1.97 \pm 0.04M_\odot$]  \cite{zhang2020gw190814} and  J0348+0432 [$2.01 \pm 0.04M_\odot$] \cite{antoniadis2013massive} brought criterion on the nature of the core matter. According to Quantum Chromodynamics [QCD], it supports the conversion of hadronic matter into deconfined quarks inside the neutron stars. This invention makes the Bodmer-Witten hypothesis quite important to the context as it already predicted for a quark-matter component in the cosmic rays detected from neutron stars \cite{bodmer1971collapsed,witten1984cosmic}. This basically strengthens the presence of Quark Stars.

The MIT bag model is the most common methodology to analyse the quark starts. The MIT bag model considers the quarks as free particles strongly confined inside hadrons as any presence of free quark is yet to be detected in particle physics. The quarks are considered to be free to move inside the hadron and are treated as free fermi gas. The confinement of quarks i.e. the bag model for the quarks can have arbitrary shapes only function of bag radius $\mathcal{R}(\theta, \phi)$ and a normal vector $n(\theta, \phi)$ out of which spherical shape is the most common natural outcomes. The parameters of the bag radius ($\mathcal{R}$) is determined by the constraint that the pressure on each quark is constant inside the bag. This theory has certain limitations as it's violates chiral symmetry and Equation of State (EoS) obtained here has no pertinent reliability on discussing systems related to quark interactions. To modify the shortcomings, Color flavour locked (CFL) matter has been proposed by few authors \cite{banerjee2021color,alford1999color} which is considered to be the ground state of QCD at extreme large densities, though the phase of the matter is not properly deterministic in this case. Asbell et. al  \cite{asbell2017oscillation} proposed a 2-component EoS of state for strange quark stars which are having mass in the range of $2 M_\odot$ by studying the non-radial oscillations emitted out from this sources. This model inspired the EoS for interacting quarks for quark stars in Einstein-Gauss-Bonnet gravity \cite{banerjee2021quark}. The derived EoS was based on simple thermodynamic bag model which considered it as homogeneous and unpaired charge neutral 3-flavor interacting quark matter in case of strange quarks.

In this paper, we will try to evaluate exact solution in $4D$ EGB-gravity for the first time that represents quark star where the internal pressure and energy density are linked by the famous MIT bag model EoS. In Sect. \ref{sec2} we will explore some aspect of $4D$ EGB gravity and its field equations for anisotropic stress tensor, Sect. \ref{sec3}  deals with physics of QCD and MIT-bag model for quark matters, a new solution is explored in Sect. \ref{sec4}. The matching conditions of the solution are provided in Sect. \ref{sec6} and the Sect. \ref{sec5} contains the physical properties of the solution for quark star. At last, finally the results and discussions are provided in Sect. \ref{sec7}.

\section{The field equations}\label{sec2}
The Lagrangian in $n-$dimensional Einstein-Gauss-Bonnet (EGB) gravity is given by
\begin{equation}
\mathcal{L}_{EGB}=\mathcal{R}+{\alpha \over n-4}~\mathcal{L}_{GB}+\mathcal{L}_{m},
\end{equation}
and corresponding action can be written as
\begin{eqnarray}
\mathcal{S}_{EGB}={1 \over 16\pi} \int d^nx \sqrt{-g} \Big(\mathcal{R}+{\alpha \over n-4}~\mathcal{L}_{GB}\Big)+\mathcal{S}_{m}. \label{e2}
\end{eqnarray}
Here $\mathcal{R}$ is the Ricci scalar, $\alpha$ the Gauss-Bonnet coupling parameter, $\mathcal{S}_{m}$ is action contributed from the matter distribution and the Gauss-Bonnet Lagrangian is given by
\begin{eqnarray}
\mathcal{L}_{GB} = \mathcal{R}^2-4\mathcal{R}_{\mu \nu} \mathcal{R}^{\mu \nu}+\mathcal{R}_{\mu \nu \kappa \lambda} \mathcal{R}^{\mu \nu \kappa \lambda}.
\end{eqnarray}
The variation of the action in \eqref{e2} with respect to the metric tensor gives
\begin{eqnarray}
\mathcal{G}_{\mu \nu} + {\alpha \over n-4}~\mathcal{H}_{\mu \nu}=8\pi \mathcal{T}_{\mu \nu}. \label{e4}
\end{eqnarray}
Here the Einstein tensor, the Lancoz tensor and the stress tensor are respectively given as
\begin{eqnarray}
\mathcal{G}_{\mu \nu} &=& \mathcal{R}_{\mu \nu} -{1 \over 2}~\mathcal{R}~g_{\mu \nu}~,\\
\mathcal{H}_{\mu \nu} &=& 2\Big(\mathcal{R} \mathcal{R}_{\mu \nu}-2\mathcal{R}_{\mu \lambda} \mathcal{R}^\lambda_\nu-2\mathcal{R}_{\mu \lambda \nu \rho} \mathcal{R}^{\lambda \rho}- \nonumber \\
&& \mathcal{R}_{\mu \alpha \beta \gamma} \mathcal{R}^{\alpha \beta \gamma}_\nu \Big)-{1 \over 2}~g_{\mu \nu} \mathcal{L}_{GB},\\
\mathcal{T}_{\mu \nu} &=& -{2 \over \sqrt{-g}}~{\delta (\sqrt{-g} ~\mathcal{S}_m) \over \delta g^{\mu \nu}}.
\end{eqnarray}
The field equation \eqref{e4} cannot give a meaningful field equation at $n=4$. However, Glavan \& Lin D. Glavan and Lin \cite{glavan2020einstein}  shown that by a re-scaling $\alpha \rightarrow \alpha / (n-4)$ and Ghosh \& Maharaj \cite{ghm}  by considering spacetimes of curvature scale $\mathcal{K}$ which are maximally symmetric determined the variation of the Gauss-Bonnet contribution as
\begin{eqnarray}
{1 \over \sqrt{-g}}~g_{\mu \lambda}~{\delta \mathcal{L}_{GB} \over \delta g_{\nu \lambda}}={\alpha(n-2)(n-3) \over 2(n-1)}~\mathcal{K}^2~\delta^\nu_\mu
\end{eqnarray}
which is clearly non-vanishing at $n=4$.

To arrive at the reduced field equations we consider an spacetime of the form in $n-$dimensions
\begin{eqnarray}
ds^2 &=&- e^\nu dt^2+e^\lambda dr^2+r^2 d\Omega^2_{n-2}. \label{e9} \label{e9}
\end{eqnarray}
Here, $d\Omega^2_{n-2}$ represents the $n-2$ dimensional surface of a unit sphere. Further, assuming stress tensor for anisotropic fluid as
\begin{eqnarray}
\mathcal{T}_{\mu \nu} = (\rho+p_\theta)u_\mu u_\nu+p_\theta g_{\mu \nu}+(p_r-p_\theta)\chi_\nu\chi^\mu,
\end{eqnarray}
where all the symbols have their usual meanings.

Now, the field equations in the limit $n\rightarrow 4$ takes the form
\begin{eqnarray}
8\pi \rho &=& {e^{-\lambda} \lambda' \over r} \left[1+{2\alpha(1-e^{-\lambda}) \over r^2}\right]+{1-e^{-\lambda} \over r^2} \nonumber \\
&& \hspace{3.5cm} \left[1-{\alpha(1-e^{-\lambda)} \over r^2} \right], \label{e11}\\
8\pi p_r &=& {e^{-\lambda} \nu' \over r} \left[1+{2\alpha(1-e^{-\lambda}) \over r^2}\right]-{1-e^{-\lambda} \over r^2} \nonumber \\
&& \hspace{3.5cm} \left[1-{\alpha(1-e^{-\lambda)} \over r^2} \right], \label{e12}\\
8\pi p_\theta &=& {e^{-\lambda} \over 4} \bigg[(2\nu''+\nu'^2)\left\{1+{4\alpha(1-e^{-\lambda}) \over r^2} \right\} + \nonumber \\
&& {2(\nu'-\lambda') \over r}~\left\{1-{2\alpha(1-e^{-\lambda}) \over r^2} \right\}-\lambda' \nu' \nonumber\\
&& \hspace{1cm}\left\{1-{8\alpha \over r^2}+{12\alpha (1-e^{-\lambda}) \over r^2} \right\} \bigg] \nonumber \\
&& \hspace{3cm}-{2\alpha(1-e^{-\lambda})^2 \over r^4}. \label{e13}
\end{eqnarray}
Since the above three field equations contain 5 unknowns, we need to assume two variables. The method of solving is considered in the next section.

\section{Quark matter and the MIT bag model} \label{sec3}
The Lagrangian in MIT bag model can be written as
\begin{eqnarray}
\mathcal{L} &=& \Big[ {i \over 2} \Big\{\overline{\psi} \gamma^\mu \partial_\mu \psi - (\partial_\mu \overline{\psi}) \gamma^\mu \psi \Big\} -\mathcal{B} \Big] \Theta(x)-{1 \over 2} \overline{\psi} \psi \Delta_s \nonumber \label{ee14}\\ 
\end{eqnarray}
where the symbols have their usual meanings. The Heaviside function and its derivative is defined as
\begin{eqnarray}
\Theta_\nu(x)=\Theta(\mathcal{R}-r) ~~~\mbox{and}~~~\partial_\mu \Theta=n_\mu \Delta_s,
\end{eqnarray}

Here $\mathcal{R}$ is the radius of the bag, inside which all the three quarks $u,~d,~s$ are able to move freely and $n_\mu$ is the unit vector normal to surface. The corresponding energy-momentum tensor of \eqref{ee14} can be determined as

\begin{eqnarray}
T^{\mu \nu} &=& \left[{\partial \mathcal{L} \over \partial (\partial_\mu \psi)}~\partial^\nu \psi+(\partial^\nu \overline{\psi})~{\partial \mathcal{L} \over \partial (\partial_\mu \overline{\psi})} \right]-\mathcal{L} \,g^{\mu \nu}
\end{eqnarray}

which will eventually become
\begin{eqnarray}
T^{\mu \nu} &=& -\mathcal{L} \,g^{\mu \nu} +{i \over 2} \big\{\overline{\psi} \gamma^\mu \partial^\nu \psi-(\partial^\nu \overline{\psi}) \gamma^\mu \psi \big\} \Theta.
\end{eqnarray}

The conservation of energy and momentum suggests
\begin{eqnarray}
\partial_\mu T^{\mu \nu} = \partial_\mu (g^{\mu \nu} \mathcal{B}\, \Theta+{1 \over 2}~g^{\mu \nu} \overline{\psi} \psi \Delta_s)=0.
\end{eqnarray}
In flat space i.e Minkowski space $\partial_\mu g^{\mu \nu}=\partial^\nu$, therefore
\begin{eqnarray}
&&\partial^\nu (\mathcal{B}\, \Theta)+{1 \over 2}~\partial^\nu (\overline{\psi} \psi \Delta_s)=0 \nonumber\\
\mbox{or} && -\mathcal{B}~n^\nu \Delta_s+{1 \over 2}~\partial^\nu (\overline{\psi} \psi ) \Delta_s=0 \nonumber\\
\mbox{or} && \mathcal{B}~n^\nu \Delta_s={1 \over 2}~\partial^\nu (\overline{\psi} \psi ) \Delta_s.
\end{eqnarray}
Here we have used $\partial^\nu \Theta = -n^\nu \Delta_s$ and $\partial^\nu \Delta_s=0$. The $\Delta_s$ function is only non-vanishing at $r=\mathcal{R}$, hence the bag constant on the surface  is 
\begin{equation}
\mathcal{B} = -{1 \over 2}~n_\nu\,\partial^\nu (\overline{\psi} \psi )
\end{equation}
as $n^\nu$, a spacelike vector. Further, the bag is considered spherically symmetric and one can write $n_\nu \partial^\nu = d/dr$, therefore 
\begin{equation}
\mathcal{B} = -{1 \over 2}{\partial (\overline{\psi} \psi) \over dr}.
\end{equation}
This means that the outward pressure due to the confined quarks on the surface of the bag is counter balancing with the inward QCD vacuum pressure $\mathcal{B}$. \\

The pressure, energy density, baryon number density and entropy at finite temperature for quarks Fermi gas of mass $m_f$ and chemical potential $\mu_f$  are
\begingroup
\small
\begin{eqnarray}
p &=& \sum_f^{u,d,s} {1 \over 3} {g_f \over 2\pi^2} \int_0^\infty \kappa {\partial E_f(\kappa) \over \partial \kappa} \Big[n(\kappa,\mu_f)+n(\kappa,-\mu_f) \Big] \kappa^2 d\kappa \nonumber \\
&& \hspace{6cm}  -\mathcal{B},\\
\rho &=& \sum_f^{u,d,s} {g_f \over 2\pi^2} \int_0^\infty E_f(\kappa) \Big[n(\kappa,\mu_f)+n(\kappa,-\mu_f) \Big] \kappa^2 d\kappa +\mathcal{B},\nonumber \\
\\
\rho_B &=& \sum_f^{u,d,s} {1 \over 3} {g_f \over 2\pi^2} \int_0^\infty \Big[n(\kappa,\mu_f)-n(\kappa,-\mu_f) \Big] \kappa^2 d\kappa,\\
S &=& \left.{\partial p \over \partial T}\right|_{V,\mu_f},
\end{eqnarray}
\endgroup
where the kinetic energy of the quarks are
\begin{equation}
E_f(\kappa)=\sqrt{m_f^2+\kappa^2}
\end{equation}
and $n(\kappa,\mu_f)$ the Fermi distribution function 
\begin{equation}
n(\kappa,\pm \mu_f) = {1 \over \exp\big[\{E_f(\kappa)\pm \mu_f\}/T\big]+1}
\end{equation}
with $T$ the temperature and $g_f=2_{spin} \otimes 3_{color}$ . Under the massless quarks approximation i.e. $m_f=0$, the pressure $p$ and the energy density $\rho$ are link by a linear equation of state (EoS)
\begin{eqnarray}
&& \hspace{0.5cm} p=\frac{1}{3} \Big(\rho-4 \mathcal{B}\Big), \label{eq28}
\end{eqnarray}
which is the famous MIT-Bag EoS.

\section{Quark star solution in MIT bag model}\label{sec4}
Following the assumption of the MIT bag EOS (\ref{eq28}), we consider the assumed MIT bag equation of state (EoS) corresponding to radial presure together with a well-behaved ansatz for the metric potential $g_{rr}$ to solve the field equations as, 
\begin{eqnarray}
&&\hspace{0.5cm} p_r ={1 \over 3}\big(\rho-4\mathcal{B}\big), \label{e14}\\
&&\hspace{0.5cm} e^\lambda = 1+ar^2+br^4, \label{e15}
\end{eqnarray}
where $\mathcal{B}$ is a bag constant, while $a$ and $b$ are arbitrary constants with dimension $l^{2}$ and $l^{-4}$, respectively. This form of the metric function is inspired due to the properties $e^{\lambda(0)}=1$ and increasing function of $r$.\\

Then the Eqs.\eqref{e11}, \eqref{e12}, \eqref{e14} and \eqref{e15} yields the metric potential $\nu$ as, 
\begingroup
\begin{eqnarray}
&& \nu(r) = {1 \over 9} \bigg[3 r^4 (-8 \pi  a \mathcal{B}+16 \pi  \alpha  b \mathcal{B}+b)+6 r^2 \big\{16 \pi  \alpha  a \mathcal{B}+ \nonumber \\
&&\hspace{0.5cm} a-3 \alpha  b-8 \mathcal{B} \left(4 \pi  \alpha ^2 b+\pi \right)\big\}+3 \log \left(a r^2+b r^4+1\right) \nonumber \\
&&\hspace{0.5cm} -16 \pi  b \mathcal{B} r^6 - \frac{3 \alpha  (32 \pi  \alpha  \mathcal{B}+3)}{\sqrt{f_1(r)}} \Big\{a^2+ a \Big(\sqrt{f_1(r)}-4 \alpha  b\Big) \nonumber \\
&&\hspace{0.5cm} +2 b \left(2 \alpha ^2 b-\alpha  \sqrt{f_1(r)}-1\right)\Big\} \log \Big\{a+2 b \left(\alpha +r^2\right) \nonumber \\
&&\hspace{0.5cm} -\sqrt{f_1(r)}\Big\}+\frac{3 \alpha  (32 \pi  \alpha \mathcal{B}+3)}{\sqrt{f_1(r)}} \Big\{a^2-a \left(4 \alpha  b+\sqrt{f_1(r)}\right) \nonumber \\
&&\hspace{0.5cm} +2 b \left(2 \alpha ^2 b+\alpha  \sqrt{f_1(r)}-1\right) \Big\}\log \Big\{a+2 b \left(\alpha +r^2\right) \nonumber \\
&&\hspace{0.5cm} +\sqrt{f_1(r)}\Big\} \bigg]+A,
\end{eqnarray}
\endgroup
where $A$ is the constant of integration and 
$f_1(r)=a^2-4 \alpha  a b+4 b \left(\alpha ^2 b-1\right)$.\\
\begin{figure} [htp!]
\includegraphics[scale=0.83]{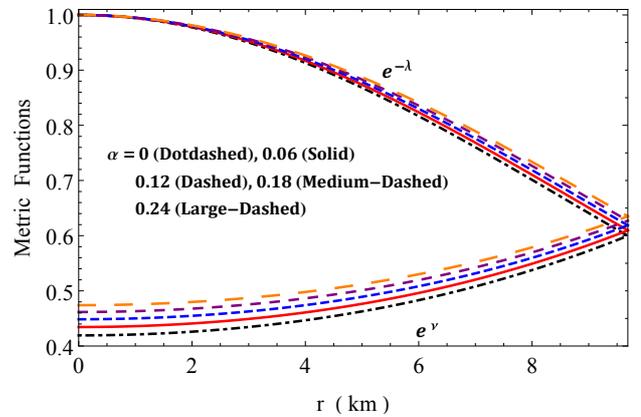}
\caption{Variation metric potentials with $r$ for $M = 1.97 M_\odot,~ R = 9.69~km,~ b = 0.000015/km^4$ and matching of boudary at the surface.}
\label{f0}
\end{figure}

Now the field equations finally give the expression for $\rho$, $p_r$, $p_t$, and $\Delta$ as, 
\begingroup
\begin{eqnarray}
\rho &=& \frac{\left(a r^2+b r^4+1\right)^{-3}}{8 \pi} \Big[\left(a+b r^2\right) \left(a r^2+b r^4+1\right)  \nonumber \\
&& \big\{-\alpha \left(a+b r^2\right)+a r^2+b r^4+1\big\}+2 \left(a+2 b r^2\right)  \nonumber \\
&& \big\{a \left(2 \alpha +r^2\right)+b \left(r^4+2 \alpha  r^2\right)+1\big\} \Big], \label{e17}\\
p_r &=& {1 \over 3} \Big(\rho-4\mathcal{B}\Big), \label{e18}\\
\Delta &=& \frac{4 \mathcal{B}}{3}-\frac{1}{24 \pi  e^{3 \lambda }} \Big[e^{\lambda } \left(a+b r^2\right) \left\{e^{\lambda }-\alpha  \left(a+b r^2\right)\right\}+2 \nonumber \\
&& \hspace{-0.7cm} \left(a+2 b r^2\right) \left(2 a \alpha +2 \alpha  b r^2+e^{\lambda }\right) \Big]+\frac{1}{8 \pi } \bigg[-\frac{2 \alpha  \left(a+b r^2\right)^2}{e^{2 \lambda }} \nonumber\\
&& \hspace{-0.7cm} +\frac{f_6(r) \big\{f_4(r)+f_5(r)\big\}}{3 e^{4 \lambda } \big\{a \left(2 \alpha +r^2\right)+b \left(r^4+2 \alpha  r^2\right)+1\big\}} -\frac{2 f_7(r)}{3 e^{3 \lambda }} \nonumber \\
&& \hspace{-0.7cm} \frac{a \left(r^2-2 \alpha \right)+b \left(r^4-2 \alpha  r^2\right)+1}{a \left(2 \alpha +r^2\right)+b \left(r^4+2 \alpha  r^2\right)+1} + \frac{1}{4 e^{\lambda }} \bigg\{\frac{4 \alpha  \left(a+b r^2\right)}{e^{\lambda }} \nonumber \\
&& \hspace{-0.7cm} +1 \bigg\} \bigg\{\frac{4 r^2 \left(f_8(r)+f_9(r)\right){}^2}{9 e^{2 \lambda } \left(a \left(2 \alpha +r^2\right)+b \left(r^4+2 \alpha  r^2\right)+1\right)^2}+{2 \over 9} \nonumber \\
&& \hspace{-0.7cm} \bigg[f_{10}(r)-\frac{12 \left(a r+2 b r^3\right)^2}{e^{2 \lambda }}+\frac{6 \left(a+6 b r^2\right)}{e^{\lambda }}-480 \pi  b B r^4 \nonumber \\
&& \hspace{-0.7cm} +\frac{48 \alpha  b^2 r^2 (32 \pi  \alpha  B+3) f_2(r)}{\sqrt{f_1(r)} \left\{a+2 b \left(\alpha +r^2\right)-\sqrt{f_1(r)}\right\}^2} -   \nonumber \\
&& \frac{12 \alpha  b (32 \pi  \alpha  B+3) f_2(r)}{\sqrt{f_1(r)} \left\{a+2 b \left(\alpha +r^2\right)-\sqrt{f_1(r)}\right\}}  \nonumber \\
&&  -\frac{48 \alpha  b^2 r^2 (32 \pi  \alpha  B+3) f_3(r)}{\sqrt{f_1(r)} \left\{a+2 b \left(\alpha +r^2\right)+\sqrt{f_1(r)}\right\}^2} \nonumber \\
&& +\frac{12 \alpha  b (32 \pi  \alpha  B+3) f_3(r)}{\sqrt{f_1(r)} \left\{a+2 b \left(\alpha +r^2\right)+\sqrt{f_1(r)}\right\}}\bigg] \bigg\} \bigg] , \label{e19}\\
p_\theta &=& \Delta+p_r,
\end{eqnarray}
\endgroup
\begin{figure} 
\includegraphics[scale=0.85]{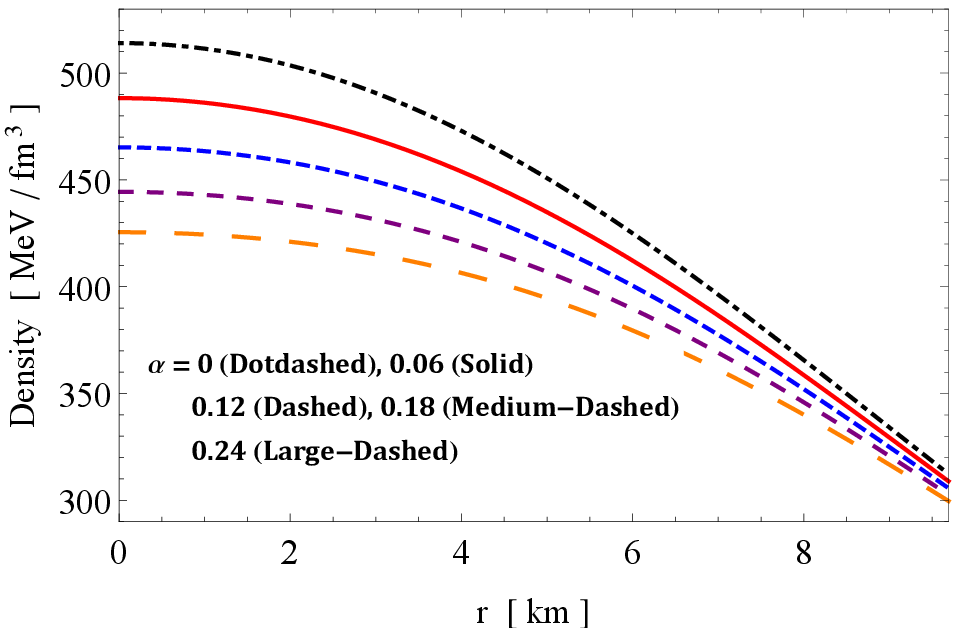}
\caption{Variation energy density with $r$ for $M = 1.97 M_\odot,~ R = 9.69~km,~ b = 0.000015/km^4$.}
\label{f1}
\end{figure}
\begin{figure} 
~~~\includegraphics[scale=0.85]{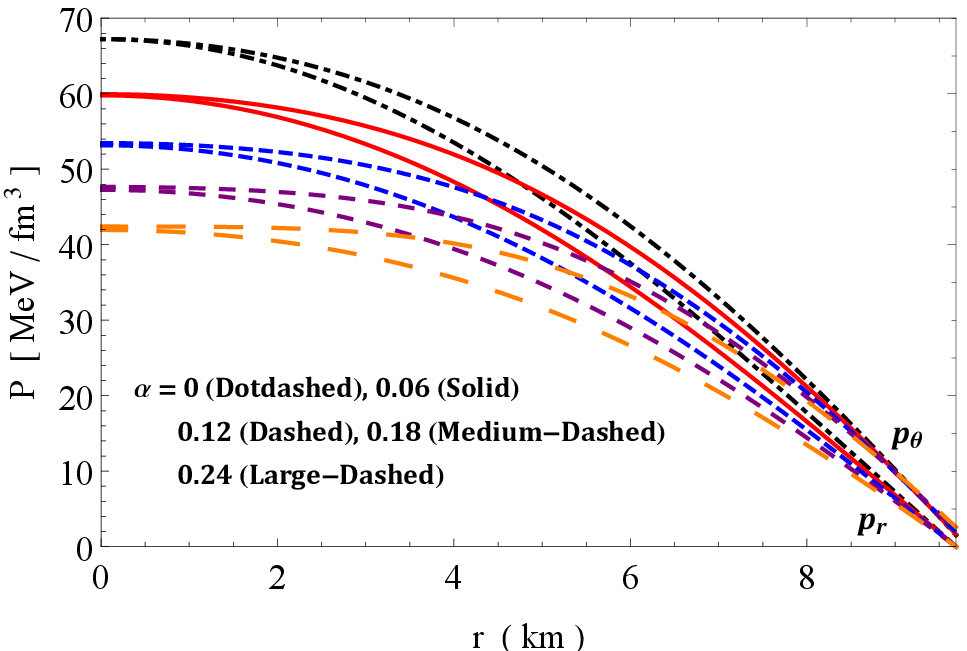}
\caption{Variation pressure with $r$ for $M = 1.97 M_\odot,~ R = 9.69~km,~ b = 0.000015/km^4$.}
\label{f2}
\end{figure}
\begin{figure} 
~~~\includegraphics[scale=0.85]{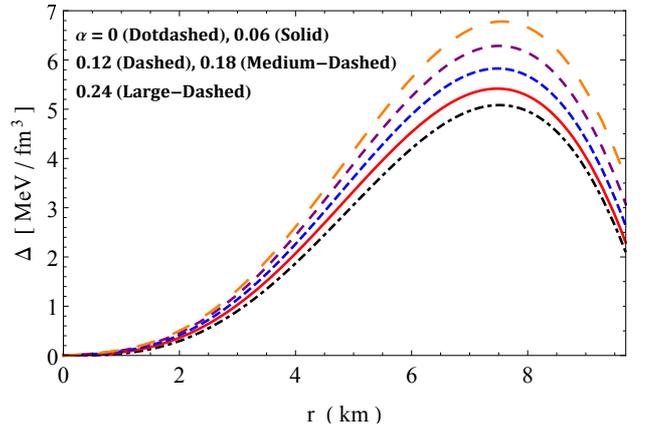}
\caption{Variation pressure anisotropy with $r$ for $M = 1.97 M_\odot,~ R = 9.69~km,~ b = 0.000015/km^4$.}
\label{f3}
\end{figure}
where,
\begingroup
\small
\begin{eqnarray*}
f_2(r) &=& a^2+a \left(\sqrt{f_1(r)}-4 \alpha  b\right)+2 b \left(2 \alpha ^2 b-\alpha  \sqrt{f_1(r)}-1\right)\\
f_3(r) &=& a^2-a \left(4 \alpha  b+\sqrt{f_1(r)}\right)+2 b \left(2 \alpha ^2 b+\alpha  \sqrt{f_1(r)}-1\right)\\
f_4(r) &=& 2 a^3 r^2 \left(\alpha +8 \pi  B r^4-r^2\right)+a^2 r^2 \big\{48 \pi  B r^2 \left(b r^4+1\right) \\
&& -6 b r^4+6 \alpha  b r^2-5\big\}+16 \pi  B \left(b r^4+1\right)^3\\
f_5(r) &=& a \big\{-6 b^2 r^8+6 \alpha  b^2 r^6+48 \pi  B \left(b r^5+r\right)^2-11 b r^4-3 \\
&& -2 \alpha  b r^2\big\}-2 b r^2 \left\{b^2 \left(r^8-\alpha  r^6\right)+b r^2 \left(\alpha +3 r^2\right)+2\right\}\\
f_6(r) &=& \left(a+2 b r^2\right) \left\{r^4 (a+4 \alpha  b)+r^2 (4 a \alpha +1)-8 \alpha +b r^6\right\}\\
f_7(r) &=& a^3 \left(\alpha  r^2-r^4\right)-a^2 \left(-3 \alpha +3 b r^6-3 \alpha  b r^4+r^2\right)-\\
&& a b r^2 \left(-8 \alpha +3 b r^6-3 \alpha  b r^4+r^2\right)+b r^2 \big\{b^2 \left(\alpha  r^6-r^8\right)\\
&& +5 \alpha  b r^2+1\big\}+8 \pi  B e^{3 \lambda }\\
\end{eqnarray*}
\endgroup
\begingroup
\small
\begin{eqnarray*}
f_8(r) &=& 2 a^3 r^2 \left(\alpha +8 \pi  B r^4-r^2\right)+a^2 r^2 \big\{48 \pi  B r^2 \left(b r^4+1\right)\\
&& -6 b r^4+6 \alpha  b r^2-5\big\}+16 \pi  B \left(b r^4+1\right)^3\\
f_9(r) &=& a \big\{-6 b^2 r^8+6 \alpha  b^2 r^6+48 \pi  B \left(b r^5+r\right)^2-11 b r^4-3\\
&& -2 \alpha  b r^2\big\}-2 b r^2 \left\{b^2 \left(r^8-\alpha  r^6\right)+b r^2 \left(\alpha +3 r^2\right)+2\right\}\\
f_{10}(r) &=& 12 \left\{16 \pi  \alpha  a B+a-3 \alpha  b-8 B \left(4 \pi  \alpha ^2 b+\pi \right)\right\}+36 r^2 \\
&& (b-8 \pi  a B+16 \pi  \alpha  b B)
\end{eqnarray*}
\endgroup
\begin{figure} 
\includegraphics[scale=0.9]{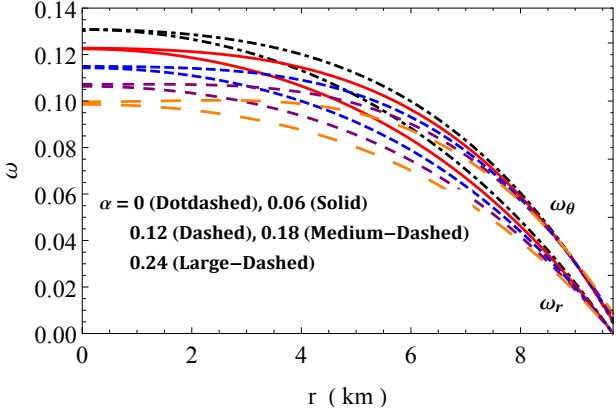}
\caption{Variation equation of state parameter with $r$ for $M = 1.97 M_\odot,~ R = 9.69~km,~ b = 0.000015/km^4$.}
\label{f4}
\end{figure}

\begin{figure} 
\includegraphics[scale=0.9]{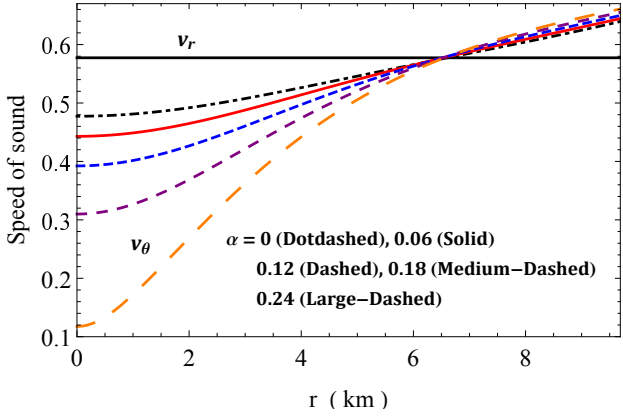}
\caption{Variation sound speed with $r$ for $M = 1.97 M_\odot,~ R = 9.69~km,~ b = 0.000015/km^4$.}
\label{f5}
\end{figure}
The rest of the physical parameters can be deduce from the density and pressures. The equation of state parameters i.e. $\omega_r=p_r/\rho$ and $\omega_\theta=p_\theta/\rho$ must be less than unity for all physical matter and unity for Zeldovich fluid ($p_z=\rho_z$).
\begin{figure} 
\includegraphics[scale=0.9]{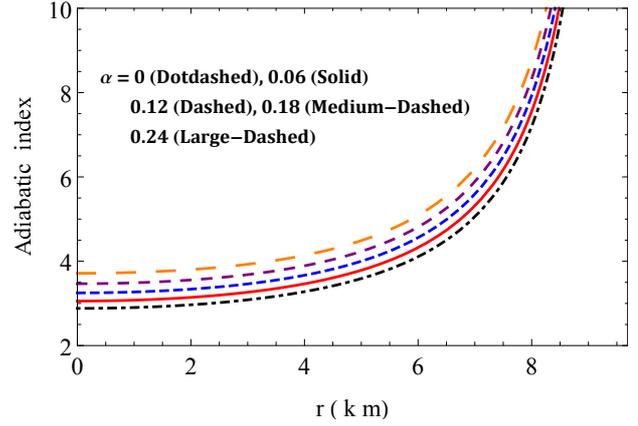}
\caption{Variation adiabatic index with $r$ for $M = 1.97 M_\odot,~ R = 9.69~km,~ b = 0.000015/km^4$.}
\label{f6}
\end{figure}
\section{Boundary conditions}\label{sec6}
To show the continuity for the interior spacetime to the external one, one need boundary matching conditions. The exterior solution is given by Glavan and Lin \cite{glavan2020einstein}  in the limit $n\rightarrow 4$ as
\begin{eqnarray}
ds^2 &=&- F(r) dt^2 + {dr^2 \over F(r)} + r^2 d\Omega_2^2,
\end{eqnarray}
where,
\begin{eqnarray}
F(r) &=& 1+{r^2 \over 32\pi \alpha} \left[1 \pm \left\{1+{128\pi \alpha M \over r^3} \right\}^{1/2} \right]. \label{e27}
\end{eqnarray}
The above exterior spacetime has no meaningful solution at short distances if $\alpha<0$, however, for $\alpha>0$ has two branches of solution. The asymptotic nature (at $r \rightarrow \infty$) of the two branches ('$-$' or '$+$') for $\alpha>0$ are
\begin{eqnarray}
F(r)\rightarrow 1-{2M \over r}~~\mbox{or}~~1+{r^2 \over 16\pi \alpha}+{2M \over r}.
\end{eqnarray}
The negative branch in \eqref{e27} asymptotically coincides with Schwarzschild exterior and hence we will choose 
\begin{eqnarray}
F(r) &=& 1+{r^2 \over 32\pi \alpha} \left[1 - \left\{1+{128\pi \alpha M \over r^3} \right\}^{1/2} \right]. \label{e29}
\end{eqnarray}
Now matching the interior and exterior spacetime at the boundary $r=R$, we get
\begin{eqnarray}
e^{\nu(R)} = e^{-\lambda(R)}=F(R).
\end{eqnarray}
The this boundary condition, we get
\begingroup
\small
\begin{eqnarray}
a &=& \frac{1-\left(b R^4+1\right) F(R)}{R^2 F(R)},\\
A &=& \log[F(R)]-{1 \over 9} \bigg[3 R^4 (-8 \pi  a \mathcal{B}+16 \pi  \alpha  b \mathcal{B}+b)+6 R^2 \big\{16 \pi    \nonumber \\
&& \hspace{-0.5cm} \alpha  a\mathcal{B}+a-3 \alpha  b-8 \mathcal{B} \left(4 \pi  \alpha ^2 b+\pi \right)\big\}+3 \log (a R^2+bR^4+1) \nonumber \\
&& -16 \pi  b \mathcal{B} R^6 - \frac{3 \alpha  (32 \pi  \alpha  \mathcal{B}+3)}{\sqrt{f_1(R)}} \Big\{a^2+ a \Big(\sqrt{f_1(R)}-4 \alpha  b\Big) \nonumber \\
&& +2 b \left(2 \alpha ^2 b-\alpha  \sqrt{f_1(R)}-1\right)\Big\} \log \Big\{a+2 b \left(\alpha +R^2\right) \nonumber \\
&& -\sqrt{f_1(R)}\Big\}+\frac{3 \alpha  (32 \pi  \alpha \mathcal{B}+3)}{\sqrt{f_1(R)}} \Big\{a^2-a \left(4 \alpha  b+\sqrt{f_1(R)}\right) \nonumber \\
&& +2 b \left(2 \alpha ^2 b+\alpha  \sqrt{f_1(R)}-1\right) \Big\}\log \Big\{a+2 b \left(\alpha +R^2\right) \nonumber \\
&& +\sqrt{f_1(R)}\Big\} \bigg]
\end{eqnarray}
\endgroup
with $f_1(R)=a^2-4 \alpha  a b+4 b \left(\alpha ^2 b-1\right)$. Further, to define the surface of the compact structure the pressure must vanishes at the surface where the two spacetime met i.e. at $r=R$ and therefore $p_r(R)=0$, we get
\begin{eqnarray}
\mathcal{B} &=& \frac{1}{32 \pi  \left(a R^2+b R^4+1\right)^3} \Big[a^3 (R^4-\alpha  R^2)+a^2 \big\{3 \alpha + \nonumber \\
&& 3 b R^6-3 \alpha  b R^4+4 R^2\big\}+a \big\{3 b^2 \left(R^8-\alpha  R^6\right)+10 b \nonumber \\
&& R^2 \left(\alpha +R^2\right)+3\big\}+b R^2 \big\{b^2 (R^8-\alpha  R^6)+b (6 R^4+ \nonumber \\
&& 7 \alpha  R^2)+5\big\} \Big]~.
\end{eqnarray}
The surface of the quark star will be use to determine the bag constant and compare with the suggested range from the quantum chromodynamical prediction. 

\section{Physical properties of the solution}\label{sec5}
The obtained solution must be tested with several physical constraints to show its physically acceptability. The variation of the metric function is shown in Fig. \ref{f0}. The meeting of the two metric functions at the surface verify the satisfaction of boundary conditions. The non-increasing nature of the energy density is shown in Fig. \ref{f1} while the decreasing nature of the pressures is shown in Fig. \ref{f2} where the radial pressure vanishes at the surface. Further, the anisotropy must be vanishing at the center for well-behaved solution. Fig. \ref{f3} shows the nature of the anisotropy, which is positive throughout the star. This implies that the anisotropic force is directed outward which helps to prevents the gravitational collapse. Now one must also verify if the equation of state (EoS) parameter is below that of Zeldovich's fluid. This is because Zeldovich's fluid is the most stiff EoS known till date with $\omega=p/\rho=1$. Fig. \ref{f4} shows that the EoS parameters along the radial as well as tangential directions are below 1 i.e. it represents a physical fluid system. Causality condition is one of the most important test to check a new theory/solution whether physical or not. This requires that the speed of sound in system must be subluminal i.e. less than that of light. For the solution the speed of sound can be found as
\begin{eqnarray}
&& \hspace{0.5cm} v^2_r = {dp_r \over d\rho}~~~~\&~~~v_\theta^2={dp_\theta \over d\rho}.
\end{eqnarray}
The variation of the speed of sound yielded from the solution is subluminal (see Fig \ref{f5}). Further, the ratio of the two specific indices i.e. $\gamma=C_p/C_v$ is an important parameter that determined a stellar system stable or not. For a Newtonian stellar fluid $\gamma>4/3$ or otherwise will initiate gravitational collapse. However, for general relativistic fluids which are generally anisotropic, deviated from the Newtonian case. For relativistic fluid the adiabatic index is given by
\begin{eqnarray}
&& \hspace{0.5cm} \gamma = {\rho+p_r \over p_r}~{dp_r \over d\rho}.
\end{eqnarray}
Here, the collapse limit depends on the nature of the anisotropy. For $p_r>p_\theta$ can even have $\gamma<4/3$ and still stable or for $p_\theta >p_r$ can be unstable even if $\gamma>4/3$. For this solution it is clear that the central values of the adiabatic index (minimum for $\alpha=0$ at 2.8) is far above 4/3 (see Fig. \ref{f6}) and hence the solution is free from gravitational collapses. 

\begin{figure} 
\includegraphics[scale=0.85]{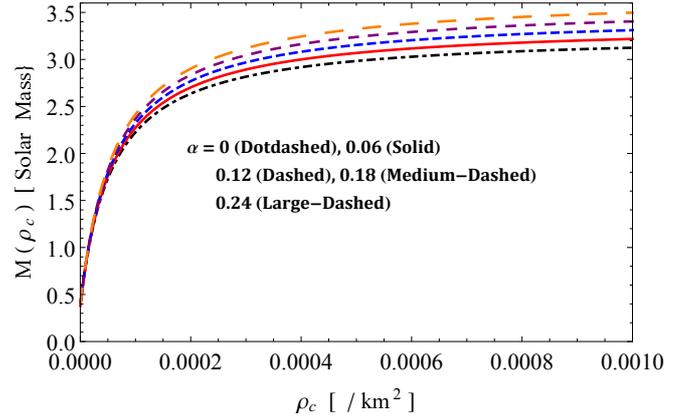}
\caption{Variation mass with central density for $b = 0.000015/km^4$.}
\label{f8}
\end{figure}

\begin{figure} 
\includegraphics[scale=0.85]{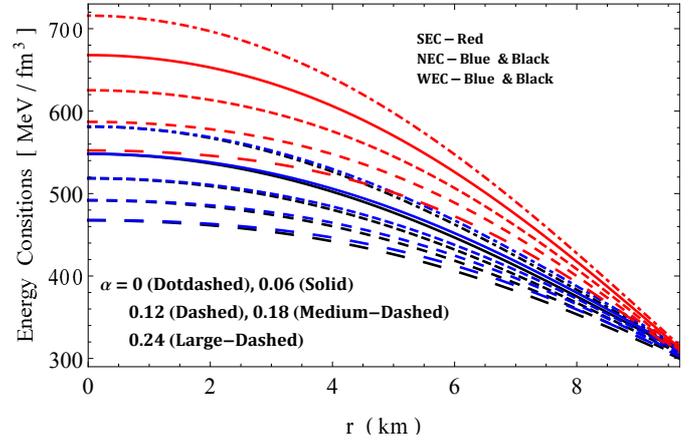}
\caption{Variation energy conditions with $r$ for $M = 1.97 M_\odot,~ R = 9.69~km,~ b = 0.000015/km^4$.}
\label{f9}
\end{figure}

All the above physical analyses clearly indicate that the solution is physically acceptable. However, one questions remains, is the system able to sustain stability under radial perturbations? This answer lies with static stability criterion. This criterion signifies that whenever the mass of a stellar fluid is an increasing function of its central density i.e. $dM/d\rho_c>0$ then system sustain stability under radial perturbations \cite{zel,har}. The mass is given by
\begin{eqnarray}
M(\rho_c) &=& \frac{R^3}{128 \pi \alpha }\bigg[
\bigg\{1+\frac{32 \pi  \alpha }{R^2} \nonumber \\
&& \hspace{1.5cm}  \left(1-\frac{1}{a R^2+bR^4+1}\right)\bigg\}^2-1\bigg]\\
a &=& \frac{\sqrt{768 \pi ^2 \alpha  \rho_c +9}-3}{6 \alpha }.
\end{eqnarray}
The variation of mass with respect to its central density in Fig. \ref{f8} shows that the solution fulfilled the static stability criterion. Finally, we must also needed to check whether the stellar fluid is physical or not using the energy conditions. For any physical fluids, the weak ($\rho \ge0,~\rho+p_i\ge 0$), null ($\rho+p_i \ge 0$), dominant ($\rho \ge |p_i|$) and strong ($\rho+\sum p_i$) energy conditions must satisfy. Figure \ref{f9} shows the satisfaction of these energy conditions. 

\begin{figure} 
\includegraphics[scale=0.9]{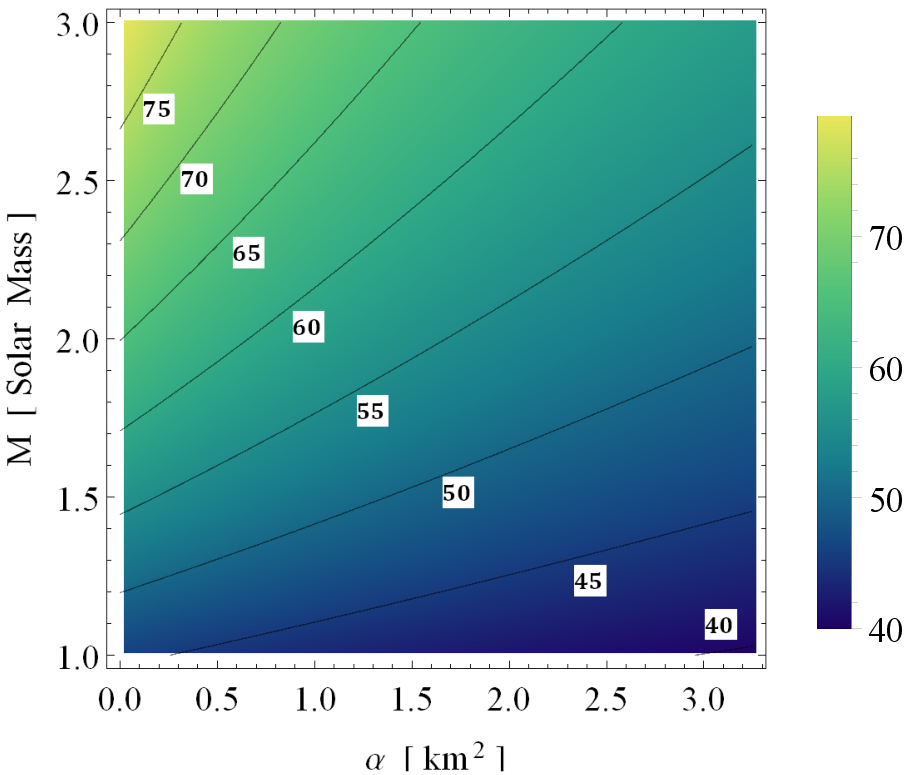}
\caption{Range of bag constant corresponding to different $M$ and $\alpha$ for $R=9.69km,~b = 0.000015/km^4$. The black lines are equi-$\mathcal{B}$ in $MeV/fm^3$.}
\label{f11}
\end{figure}

\begin{figure} 
\includegraphics[scale=0.9]{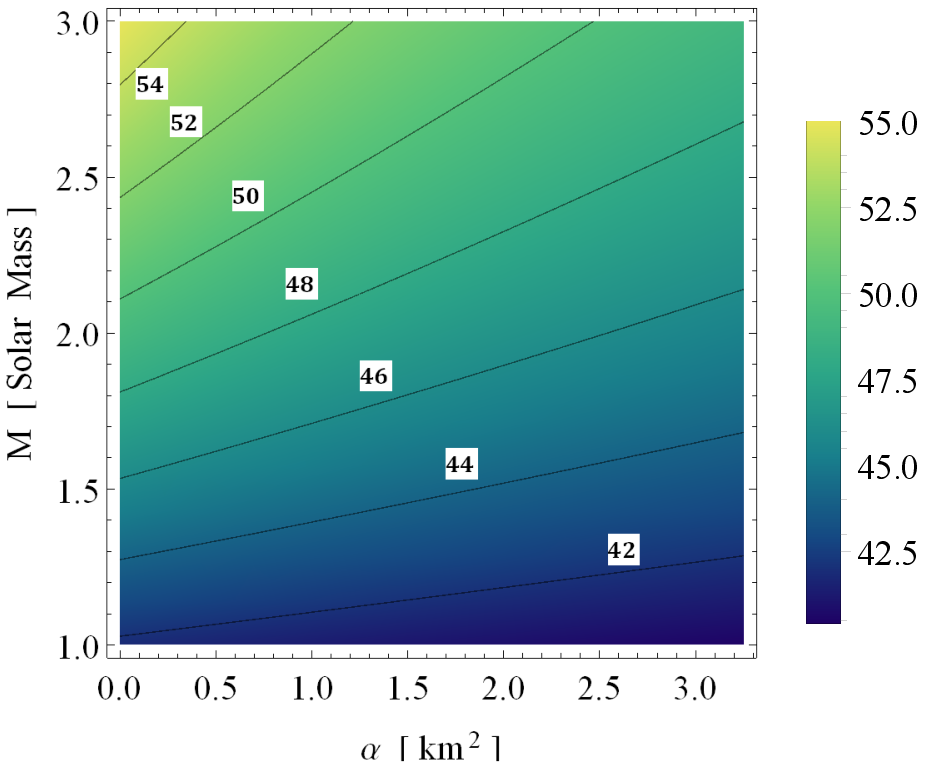}
\caption{Range of bag constant corresponding to different $M$ and $\alpha$ for $R=12km,~b = 0.000015/km^4$.The black lines are equi-$\mathcal{B}$ in $MeV/fm^3$.}
\label{f12}
\end{figure}

\begin{figure} 
\includegraphics[scale=0.85]{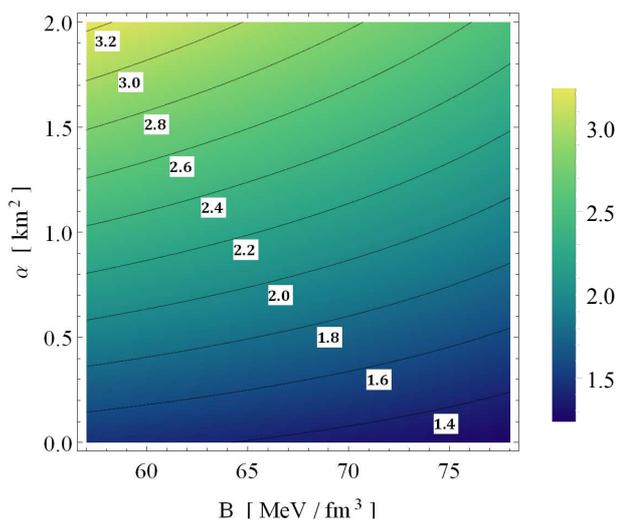}
\caption{$\mathcal{B}-\alpha$ plane for equi-$M/M_\odot$ for $R=9.69km,~b = 0.000015/km^4$.}
\label{f13}
\end{figure}


\begin{table*}
\caption{\label{tab} Values of all the parameters corresponding to different values of $\alpha$ and for $M=1.97M_\odot,~R=9.69km$.}
\centering
\begin{tabular}{llllllllllp{0.65in}}
\hline
$\alpha$ & $b\times$ 10$^{-5}$ & $\rho_s\times$ 10$^{14}$ & $p_c\times$ 10$^{35}$ & $\mathcal{B}$ & $z_s$\\
$(km^{2})$ & $(km^{-4})$ &($g/cm^3$)& ($dyne/cm^2$) & $(MeV/fm^3)$ & \\
\hline
0 & $1.5 $ & 5.576 & 1.075 & 89.85 & 0.290\\
0.06 & $1.5$ & 5.515 & 0.960 & 89.51 & 0.280\\
0.12 & $1.5$ & 5.455 & 0.854 & 89.15 & 0.270\\
0.18 & $1.5$ & 5.406 & 0.761 & 88.79 & 0.263\\
0.24 & $1.5 $ & 5.346 & 0.676 & 88.44 & 0.255\\
\hline
\end{tabular}
\end{table*}
On the other hand, the predicted range of $\mathcal{B}$ for $R=9.69\,km$ and $12\,km$ with the same mass range is shown in Figs. \ref{f11} and \ref{f12}. These figures show the ranges of $\mathcal{B}$ by varying the mass $M$ and Gauss-Bonnet coupling constant $\alpha$. As can be seen that for same mass and radius the bag constant decreases with increase in coupling constant $\alpha$. Also, a high mass quark star configuration is possible for low $\alpha$ and high $\mathcal{B}$ or high $\alpha$ and low $\mathcal{B}$. Further, the ranges of bag constant estimated for $0\le \alpha \le 3$ and $1M_\odot \le M \le 3M_\odot$ for 9.69 km radius is $45 MeV/fm^3 \le \mathcal{B} \le 80 MeV/fm^3$ and for 12 km radius is  $40 MeV/fm^3 \le \mathcal{B} \le 56 MeV/fm^3$. The predicted range of bag constant lies between $57MeV/fm^3$ and $92MeV/fm^3$ \cite{bla}. This clear means that, for example, a quark star of mass $2M_\odot$ with radius 12 km (See Fig. \ref{f12}) is highly unlikely because in such case the bag constant should be in range $46MeV/fm^3 < \mathcal{B} < 50MeV/fm^3$ which well below the predicted range. The $\mathcal{B}-\alpha$ plan with equi-mass lines are also plotted in Fig. \ref{f13} for the ranges $0\le \alpha \le 2$ and $57MeV/fm^3 \le \mathcal{B} \le 78MeV/fm^3$. This plot clearly shows that the higher masses of a quark star configuration is possible for low bag constant and higher Gauss-Bonnet coupling constant. This also implies that reducing the bag constant make the EoS stiff with the increase of $\alpha$-coupling. Table \ref{tab} shows how the physical parameters such as surface density, central pressure, surface redshift and bag constant varies with the Gauss-Bonnet coupling strength for a particle stellar system of mass $1.97M_\odot$ and radius 9.69 $km$.

\section{Results and discussions} \label{sec7}
We have successfully discover the first exact solution describing quark stars in 4-$D$ EGB gravity. This solution match with the exterior solution smoothly at the surface of the star $r=R$ (Fig. \ref{f0}), which has non-increasing energy density \& pressure (Figs. \ref{f1} \& \ref{f2}) as well as vanishing anisotropy at $r=0$ (Fig. \ref{f3}). For this solution, the EoS parameters are less than unity (Fig. \ref{f4}) and satisfy the causality condition (Fig. \ref{f5}). Moreover, the solution has been also tested with stringent physical constraints like Bondi criterion, static stability criterion and energy conditions. As per the central values of adiabatic index (see Fig. \ref{f6}) and the static stability criterion (see Fig. \ref{f8}), one can clearly see the $\gamma_c$ is more and more stable density range for  higher coupling constant $\alpha$. This only implies that the quark star configuration is more stable with larger coupling strength i.e. when the Gauss-Bonnet term is contribution more into the field equations. Moreover, the solution also fulfilled all the energy conditions which was required for any physical system (Fig. \ref{f9}). The interior solution was also matched with the Schwarzschild-like exterior spacetime. One of these matching conditions was also used to predicts the range of bag constant $\mathcal{B}$ without any concepts of microscopic QCD calculations. The $\alpha-M$ contour graphs in Figs. \ref{f11} and \ref{f12} shows that which configurations of quark stars are physically motivated keeping in mind the predicted range of $\mathcal{B}$ as per QCD calculations. The $\mathcal{B}-\alpha$ plot with equi-mass contours clearly indicate that quark star mass gets heavier with decreasing bag constant and increasing $\alpha$ coupling. The present work has three significant results (i) it might have astrophysical interest in future, (ii) a way of predicting the range of bag constant in high energy density regime and (iii) can determine the mass of quark star configuration for wide ranges of $\alpha$ and $\mathcal{B}$.

\textcolor{blue}{To see if the 4D EGB gravity reduces to GR under the condition $\alpha \rightarrow 0$, one has to see its vacuum solution. The vacuum solution in \eqref{e27} at $\alpha \rightarrow 0$ reduces to
\begin{eqnarray}
F(r)  \approx 1+{r^2 \over 32\pi \alpha} \Big[1\pm \Big\{1+{1 \over 2} \, {128\pi \alpha M \over r^3} + \text{O}(\alpha^2)\Big\} \Big].
\end{eqnarray}
Hence, the negative branch of the above function reduces to
\begin{eqnarray}
F(r) \rightarrow 1-{2M \over r},
\end{eqnarray}
which is again the Schwarzschild vacuum solution. However, the positive branch reduces to
\begin{eqnarray}
F(r) \approx  1+{r^2 \over 32\pi \alpha} \Big[2+{1 \over 2} \, {128\pi \alpha M \over r^3} + \text{O}(\alpha^2) \Big],
\end{eqnarray}
which is undefined at $\alpha \rightarrow 0$. Therefore, the negative branch of the vacuum solution coincides with the Schwarzschild's vacuum solution under $\alpha \rightarrow 0$ and $r \rightarrow \infty$ conditions while the positive branch doesn't coincide under these two conditions. Therefore, only the negative branch makes it more physical and the 4D EGB gravity can reduce to GR. However, both the negative and positive branches are solutions of 4D EGB gravity hence it cannot completely reduces to GR at $\alpha \rightarrow 0$. Further, since the positive branch blows-up both at $r \rightarrow \infty$ and $\alpha \rightarrow 0$, this solution is physically meaningless at least in 4D-gravity.}

\begin{acknowledgements}
Authors are grateful to Prof. Sushant Ghosh, Centre for Theoretical Physics, Jamia Millia Islamia, New Delhi for useful suggestions on improving the contents of the paper. The author S. K. Maurya acknowledge that this work is carried out under TRC project-BFP/RGP/CBS/19/099 of the Sultanate of Oman. The author also acknowledges to University of Nizwa administration for their continuing support
and encouragements.  FR would like to thank the authorities of the Inter-University Centre for Astronomy and Astrophysics, Pune, India for providing research facilities.
This work is a part of the project submitted by FR in DST-SERB, Govt. of India.
\end{acknowledgements}


\begin{thebibliography}{99}

\bibitem{ghosh2021phase} S. G. Ghosh, D. V. Singh, R. Kumar,  S. D. Maharaj, Ann.  Phys.  \textbf{424}, 168347 (2021)
\bibitem{lovelock1971einstein} D. Lovelock. J. Math. Phys. \textbf{12}, 498-501(1971)
\bibitem{lovelock1972four} D. Lovelock.  J. Math. Phys. \textbf{13}, 874-876 (1972)
\bibitem{lanczos1938remarkable} C. Lanczos.  Ann. Math. \textbf{39}, 842-850 (1938)
\bibitem{zwiebach1985curvature} B. Zwiebach. Phys. Lett. B, \textbf{156}, 315-317 (1985)
\bibitem{wiltshire1986spherically} D. L.  Wiltshire. Phys. Lett. B, \textbf{169}, 36-40 (1986).
\bibitem{wheeler1986symmetric} J. T.  Wheeler.  Nuc. Phys. B, \textbf{273}, 732-748 (1986)
\bibitem{boulware1985string} D. G. Boulware, S. Deser.  Phys. Rev. Lett.  \textbf{55}, 2656 (1985)
\bibitem{glavan2020einstein} D. Glavan , C. Lin.  Phys. Rev. Lett. \textbf{124}, 081301 (2020)
\bibitem{mardones1991lovelock} A. Mardones , J. Zanelli.  Class. Quan. Grav. \textbf{8}, 1545 (1991)
\bibitem{torii2008n+} T. Torii , H. Shinkai.  Phys. Rev. D
\textbf{78}, 084037 (2008)
\bibitem{woodard2015theorem} R. P. Woodard.  arXiv:1506.02210, (2015)
\bibitem{tomozawa2011quantum} Y. Tomozawa.  arXiv:1107.1424, (2011)
\bibitem{gurses2020there} M. Gurses, T.  C¸ . Sisman,  B. Tekin, Eur. Phys. J. C \textbf{80}, 1-6 (2020)
\bibitem{ghosh2020generating} S. G. Ghosh, R. Kumar.  Class. Quan. Grav.  \textbf{37}, 245008 (2020)
\bibitem{konoplya2020stability} R. A.  Konoplya , A. Zhidenko.  Phys. Dark Univ \textbf{30}, 100697 (2020)
\bibitem{kumar2020hayward} A. Kumar , S. G. Ghosh.  arXiv:2004.01131, (2020)
\bibitem{kumar2020bardeen} A. Kumar , R. Kumar. arXiv:2003.13104, (2020)
\bibitem{zhang2020superradiance} C. Zhang, S. Zhang, P. Li,  M. Guo. JHEP \textbf{2020}, 1-19 (2020)
\bibitem{mansoori2021thermodynamic} S. A. H. Mansoori.  Phys. Dark Univ. \textbf{31}, 100776 (2021)
\bibitem{mishra2020quasinormal} A. K. Mishra.  Gen. Rel. Grav. \textbf{52}, 1-18 (2020)
\bibitem{zhang2020spinning} Y. Zhang, S. Wei,  Y.  Liu.  Universe, \textbf{6}, 103 (2020)
\bibitem{zhang2020greybody} C. Zhang, P. Li,  M. Guo. The Eur. Phys. J. C, \textbf{80}, 1-9 (2020)
\bibitem{kimet2020wormholes} J. Kimet, B. Ayan, S. G. Ghosh. The Eur. Phys. J. C,  \textbf{80}, 698 (2020)
\bibitem{zhang2020gw190814} N. Zhang , B. Li. Astrophys. J. \textbf{902}, 38 (2020)
\bibitem{antoniadis2013massive}  J.   Antoniadis et  al., Science, \textbf{340}, 6131 (2013)
\bibitem{bodmer1971collapsed} A. R. Bodmer. Phys. Rev. D  \textbf{4}, 1601  (1971)
\bibitem{witten1984cosmic} E. Witten. Phys. Rev. D, \textbf{30}, 272  (1984)
\bibitem{banerjee2021color} A. Banerjee, K. N. Singh. Phys. Dark Univ. \textbf{31}, 100792 (2021)
\bibitem{alford1999color} M.  Alford,  K.  Rajagopal,  F.  Wilczek, Nuclear Physics B \textbf{537}, 443-458 (1999)
\bibitem{asbell2017oscillation} J.  Asbell, P.  Jaikumar, Journal of Physics: Conference Series \textbf{861}, 012029 (2017)
\bibitem{banerjee2021quark} A. Banerjee, T. Tangphati, D. Samart, P. Channuie.  Astrophys. J. \textbf{906}, 114 (2021)
\bibitem{ghm} S. G. Ghosh, S. D. Maharaj, Phys. Dark Univ. \textbf{30}, 100687 (2020)
\bibitem{zel} Y. B. Zeldovich, I. D. Novikov, Relativistic Astrophysics, Vol. 1: Stars and Relativity (Univ. of Chicago Press, 1971).
\bibitem{har} B. K. Harrison et al., Gravitational Theory and Gravitational Collapse (Univ. of Chicago Press, 1965)
\bibitem{bla} D. Blaschke, N. Chamel, Astrophys. Space Sci. Libr. \textbf{457}, 337 (2018)

\end{thebibliography}

\end{document}